\documentclass[epj,twocolumn]{webofc} 
\usepackage[varg]{txfonts}   

\usepackage{xspace}
\usepackage{xcolor}
\graphicspath{{plots/}}

\newcommand{\sibyll}[1]{{\sc Sibyll#1}}

\def\nmu{\ensuremath{N_\mu}\xspace}

\begin{document}
\title{Probing the $\pi^0$ spectrum at high-$x$ in proton-Air interactions at ultra-high energies}
%
%

\author{\firstname{Lorenzo} \lastname{Cazon}\inst{1} \and
  \firstname{Ruben} \lastname{Concei\c{c}\~{a}o}\inst{1,2} \and
  \firstname{Miguel Alexandre} \lastname{Martins}\inst{1,2} \and
  \firstname{Felix} \lastname{Riehn}\inst{1}\fnsep\thanks{\email{friehn@lip.pt}}
}

\institute{Laborat\'{o}rio de Instrumenta\c{c}\~{a}o e F\'{i}sica Experimental de Part\'{i}culas (LIP) - Lisbon, Av.\ Prof.\ Gama Pinto 2, 1649-003 Lisbon, Portugal
  \and
  Instituto Superior T\'ecnico (IST), Universidade de Lisboa, Av.\ Rovisco Pais 1, 1049-001 Lisbon, Portugal
}

\abstract{%
The average number of muons in air showers and its connection with shower development has been studied extensively in the past. With the upcoming detector upgrades, UHECR observatories will be able to probe higher moments of the distribution of the number of muons. Here a study of the physics of the fluctuations of the muon content is presented. In addition to proving that the fluctuations must be dominated by the first interactions, we show that low-$N_{\mu}$ tail of the shower-to-shower distribution of the number of muons is determined by the high-$x_{\rm L}$ region of the production cross-section of neutral pions in the first interaction.
}
\maketitle
\section{Introduction}



Due to the missmatch between the predicted and observed number of muons at ground~\cite{Aab:2014pza,Aab:2016hkv,Abbasi:2018fkz,Aartsen:2017upd,uhecrAuger}, much of the efforts to understand muon production in extensive air showers has been focused on the study of the average muon content~\cite{Matthews:2005sd,Maris:2009uc,Ulrich:2010rg}. The experimental situation is summarized in the report of the working group on hadronic interactions~\cite{WHISP}. These studies suggest that the average muon number very much depends on low-energy interactions in the shower. In a recent paper~\cite{Cazon:2018gww}, it was shown that, in contrast to the average, the relative fluctuations of the number of muons, $\sigma(N_{\mu})/\langle N_{\mu} \rangle$, to a large degree are determined by the first interaction. In fact, it was shown that the very shape of the shower-to-shower distribution of the number of muons is determined mostly by the fluctuations of the \textit{hadronic energy}, $E_{\rm had}$, i.e.\ the energy carried by the particles that are likely to undergo another hadronic interaction. In the following we show that these fluctuations of hadronic energy, are linked to the energy spectrum of neutral pions in the first interaction. Ultimately, it is shown that the tail of the shower-to-shower distribution of the number of muons is determined by the production cross-section of neutral pions in the first interaction. 

\section{Shower-to-shower distribution of the muon content}
\label{sec:nmu-distr}

\begin{figure}
  \centering
  \includegraphics[width=0.98\columnwidth,clip]{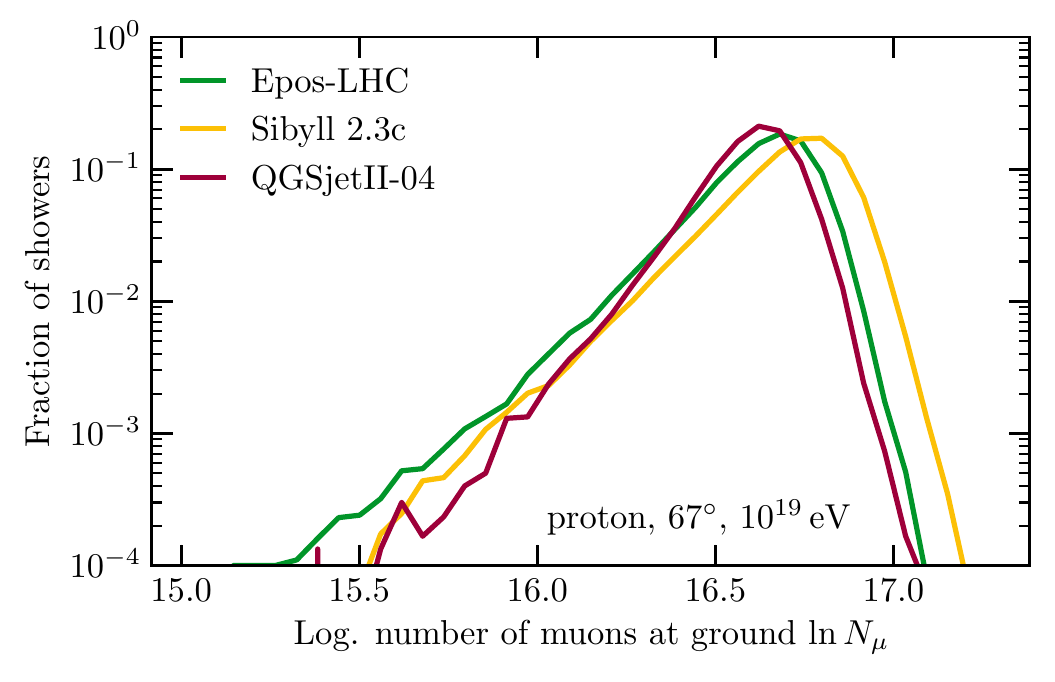}
  \caption{Distribution of the number of muons at ground in extensive air showers induced by protons with an energy of $10^{19}\,$eV. }
  \label{fig-nmu-distr}
\end{figure}

The shower-to-shower distribution of the number of muons arriving at ground from simulated air showers that were initiated by protons with a primary energy of $10^{19}\,$eV is shown in Fig.~\ref{fig-nmu-distr}. The average logarithm of the number of muons at ground, $\ln N_{\mu}$, is around 16.5 (15 million muons above $1\,$GeV). From the maximum, the distribution falls off quickly towards higher muon numbers. On the side of low muon numbers the distribution exhibits a long tail. How does this distribution arise? Which processes shape the distribution?

Muons are produced in EAS by the decay of hadrons. This means muon production is very much linked to the hadronic cascade and hadronic multiparticle production. In contrast to electromagnetic (em.) cascades, which at high energies mostly involve two processes, electron-pair creation and bremsstrahlung, the hadronic cascade (equations) cannot be solved analytically. However, as our goal is explaining the fluctuations of the number of muons, the cascade equations and their solutions, which describe average quantities only, are not the ideal tool here. The alternative approach, namely to fully simulate the development of air showers in all microscopic detail on a computer, while certainly possible~\cite{Heck98a}, due to the large number of degrees of freedom, lacks the necessary transparency to help explaining how the final state, e.g.\ Fig.~\ref{fig-nmu-distr}, arises. A successful strategy is to adopt elements from both approaches~\cite{Drescher:2002cr,Bergmann:2006yz}.

\begin{figure}
  \centering
  \includegraphics[width=0.8\columnwidth,clip]{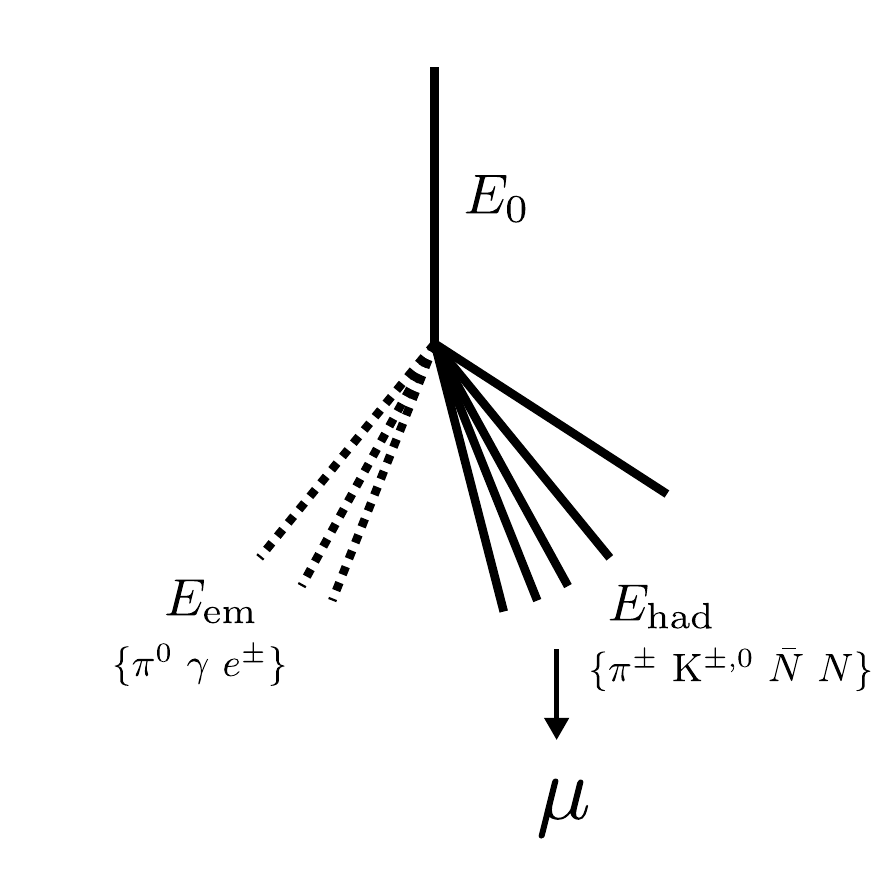}
  \caption{Energy flow in the first interaction into the em.\ and hadronic component of the shower.}
  \label{fig-first-int}
\end{figure}

In an effort to reduce the complexity of hadronic interactions we start by neglecting all differences between hadrons, except for the distinction between hadrons that instantaneously decay into electrons and/or photons and those that do not~\cite{Matthews:2005sd}. The former group will be referred to as the \textit{electromagnetic} component, the latter as \textit{hadronic} component~\footnote{More precisely, particles in the hadronic component are \textit{reinteracting} hadrons. Charmed hadrons which immediately decay, possibly into muons, do not count, as they do not contribute to the cascade.}. Approximate solutions to the cascade equations~\cite{Gaisser:2016uoy}, full simulations~\cite{Engel:2011zzb} and measurements of air showers~\cite{Aab:2014pza} show that the average number of muons in a hadron cascade as a function of the energy can be described by an expression of the form,
\begin{equation}
  \langle N_{\mu}(E) \rangle ~\sim~ E^{\beta} \ .
  \label{eq-avg-nmu}
\end{equation}
Microscopically, the exponent $\beta \approx 0.93$ is related to the balance between the production of em.\ and hadronic particles~\cite{Matthews:2005sd}.

\begin{figure}
  \centering
  \includegraphics[width=0.98\columnwidth,clip]{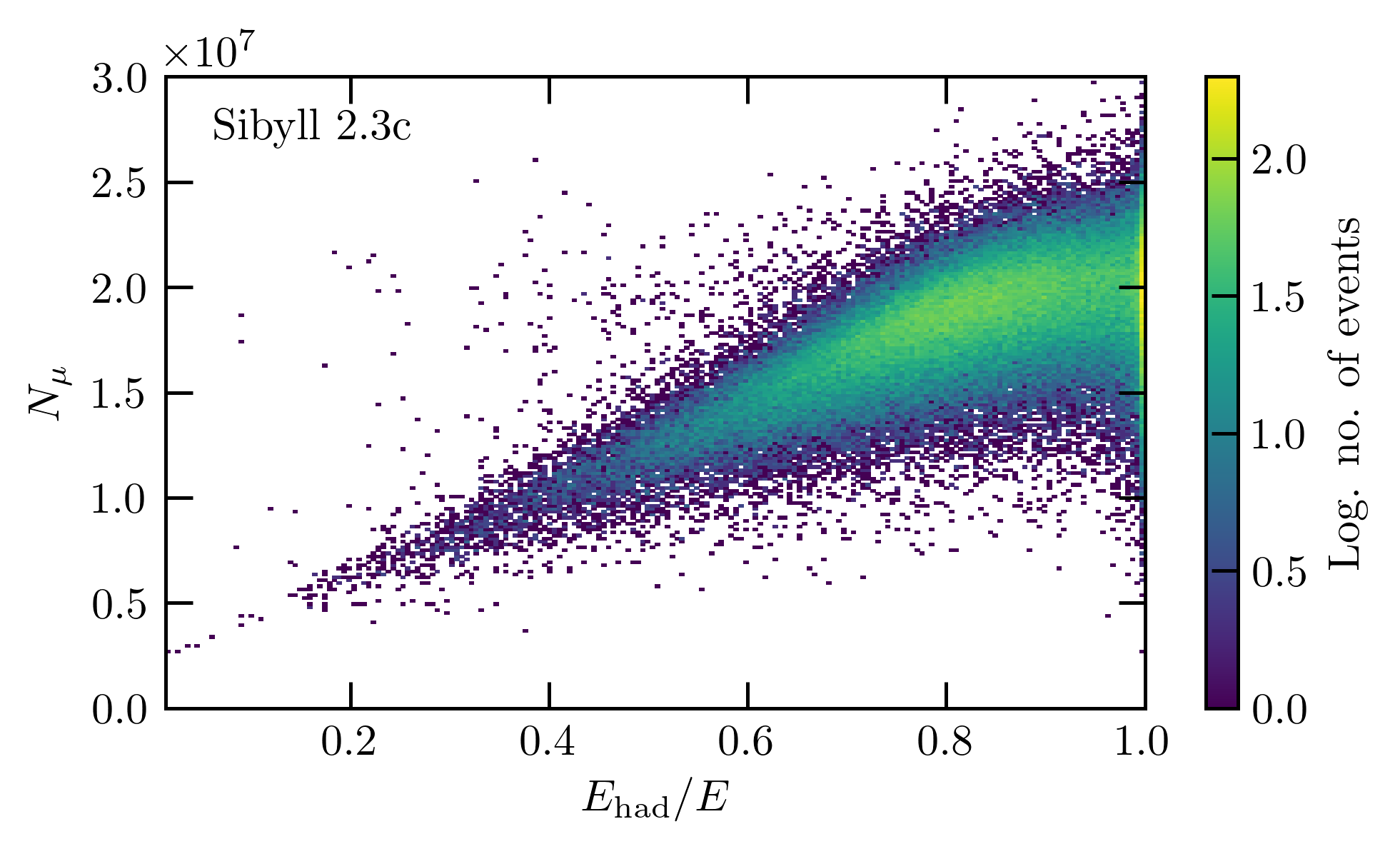}
  \caption{Distribution of the hadronic energy and the number of muons at ground in proton induced air showers with a primary energy of $10^{19}\,$eV.}
  \label{fig-nmu-alpha}
\end{figure}

\begin{figure}
  \centering
  \includegraphics[width=0.98\columnwidth,clip]{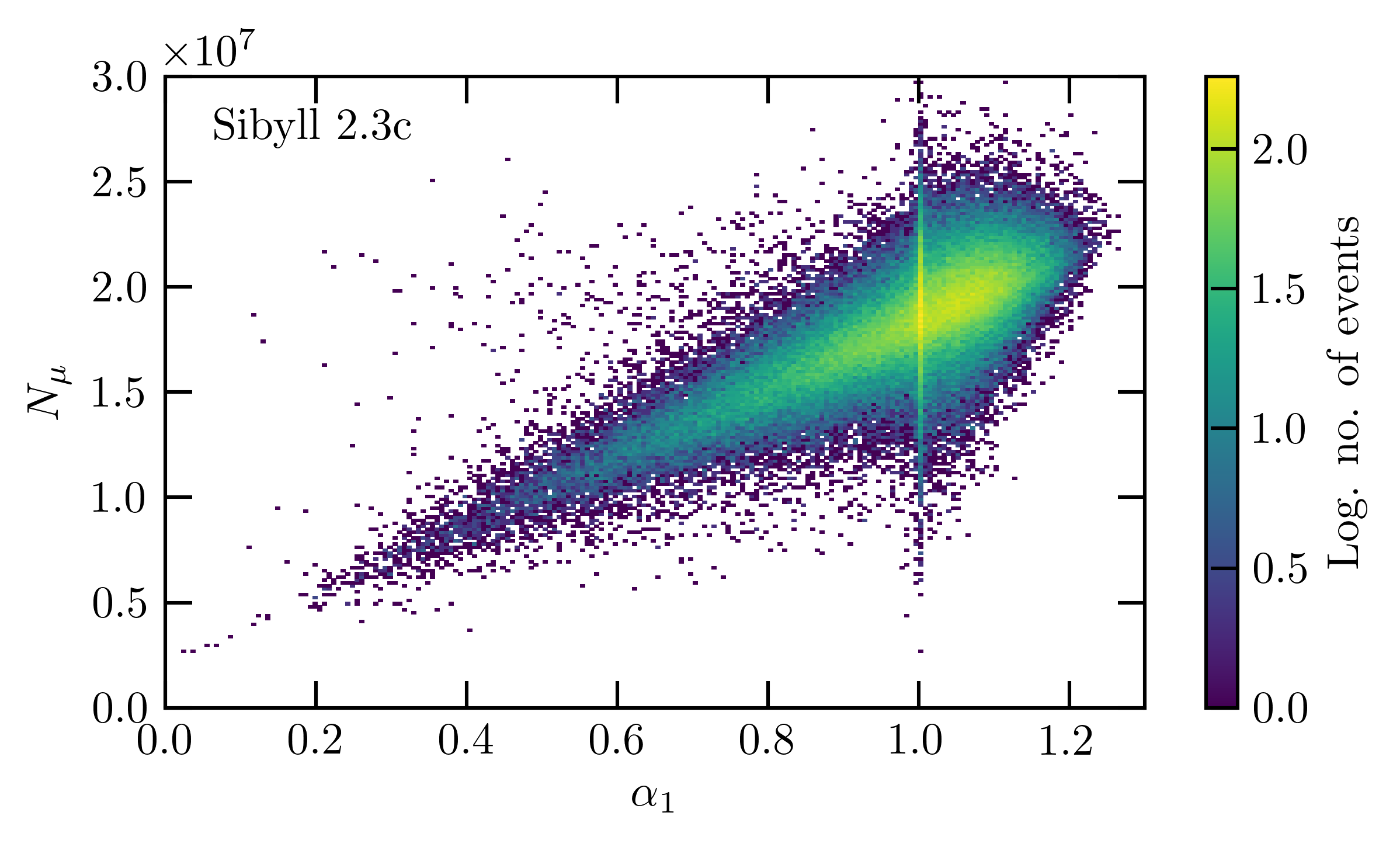}
  \caption{Distribution of $\alpha_1$ and the number of muons at ground in proton induced air showers with a primary energy of $10^{19}\,$eV.}
  \label{fig-nmu-alphaGamma}
\end{figure}

As stated in the beginning, muons are essentially produced in the hadronic cascade\footnote{16\% from $\mu$ pair production and photo-nuclear interactions.}. Combined with the idea that the average muon yield of the cascade is determined by the energy (Eq.~\eqref{eq-avg-nmu}), we can expect that the number of muons produced in the shower following the first interaction is proportional to the available energy $E_{\rm had}$ (illustrated in Fig.~\ref{fig-first-int}). Taking full air shower simulations and looking at hadronic multiparticle production in the first interaction~\cite{Riehn:2017mfm,Pierog:2013ria,Ostapchenko:2010vb} we find that the hadronic energy, $E_{\rm had}$, is correlated with the number of muons at ground in the region of low numbers of muons (see Fig.~\ref{fig-nmu-alpha}). In the case that most of the energy remains in the hadronic component, corresponding to air showers with a large number muons at ground (see Fig.~\ref{fig-nmu-alpha} in the range where $E_{\rm had}/E_0= 0.8$-$1$), the correlation between the number of muons and $E_{\rm had}$ is small. Taking the multiplicity of hadrons that are produced in the first interaction into account, this region can be described as well and the correlation can be increased further~\cite{Cazon:2018gww}. Comparing the size of the fluctuations, we find that the variance of the hadronic energy and multiplicity in the first interaction accounts for 80\% of the total variance\footnote{Correlation between the first interaction and the rest of the shower is small ($0.02$)}. Specifically, this is achieved by defining
\begin{equation}
  \alpha_1~=~ \sum_{i=1}^{m} x_{\mathrm{L},i}^{\beta} \ ,
\end{equation}
where $\beta$ is the slope parameter of the average number of muons defined previously (see Eq.~\eqref{eq-avg-nmu}), $m$ is the number of particles in the hadronic component (multiplicity) and $x_{\mathrm{L},i}~\equiv~E_{i}/E_{0}$ is the fraction of the primary energy carried by particle $i$. The variable $\alpha_1$ effectively interpolates between pure hadronic energy, $E_{\rm had}/E_0 = \sum_{i}^{m}x_{\mathrm{L},i}$,  and the hadronic multiplicity, $m$. The correlation between $\alpha_1$ and the number of muons at ground is shown in Fig.~\ref{fig-nmu-alphaGamma}.

The overall picture, however, remains valid: fluctuations of the number of muons are energy fluctuations in the first interaction.


\section{Origin of the fluctuations of $E_{\rm had}$}

\begin{figure}[h]
  \centering
  \includegraphics[width=\columnwidth,clip]{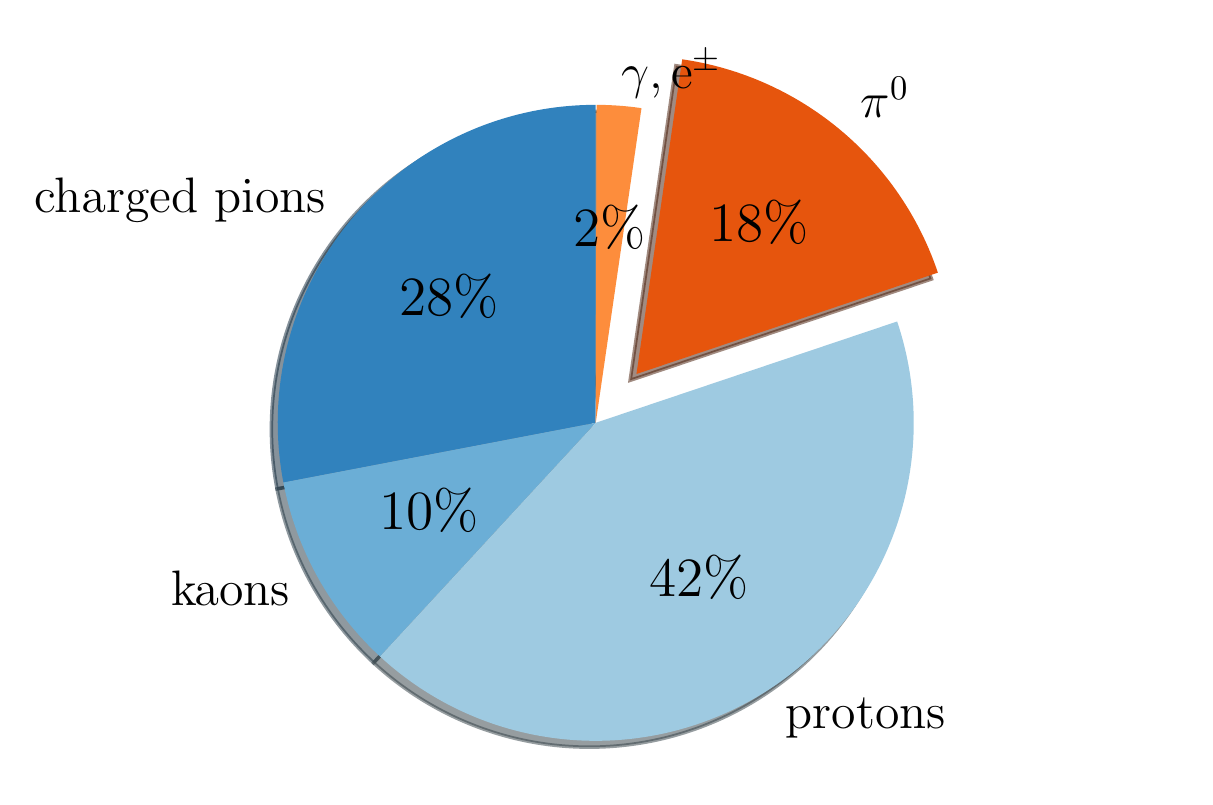}
  \caption{Distribution of the energy among different groups of particles in the first interaction. Numbers are in percent. Particles in the electromagnetic component are shown in red, particles in the hadronic component in blue. The energy carried by neutral pions is shown in the slightly removed slice. The component labeled 'protons' corresponds to p,n,$\Lambda$ and their antiparticles.}
  \label{fig-en-share}
\end{figure}

Having established the connection between energy fluctuations and the number of muons, the question becomes: what drives the fluctuations of the energy? To answer this question it is easiest to switch components. Clearly, $E_{\rm had}~=~ E_0 - E_{\rm em}$ and $\sigma(E_{\rm had})~=~\sigma(E_{\rm em})$, so the electromagnetic and hadronic components are equivalent. However, the em.\ component is dominated by neutral pions alone, whereas the hadronic component contains equal contributions from pions and protons (baryons) (see Fig.~\ref{fig-en-share}).

This means the fluctuations of $E_{\rm had}$ are related to the energy distribution of neutral pions. The region of the tail is populated by events where most of the energy is taken by a $\pi^0$ and the remaining energy is shared by other hadrons. In the language of hadronic multiparticle production, the tail of the distribution of hadronic energy, and by extension the distribution of the number of muons is linked to the \textit{inclusive production cross-section} of neutral pions.

\begin{figure}[h]
  \centering
  \includegraphics[width=0.98\columnwidth,clip]{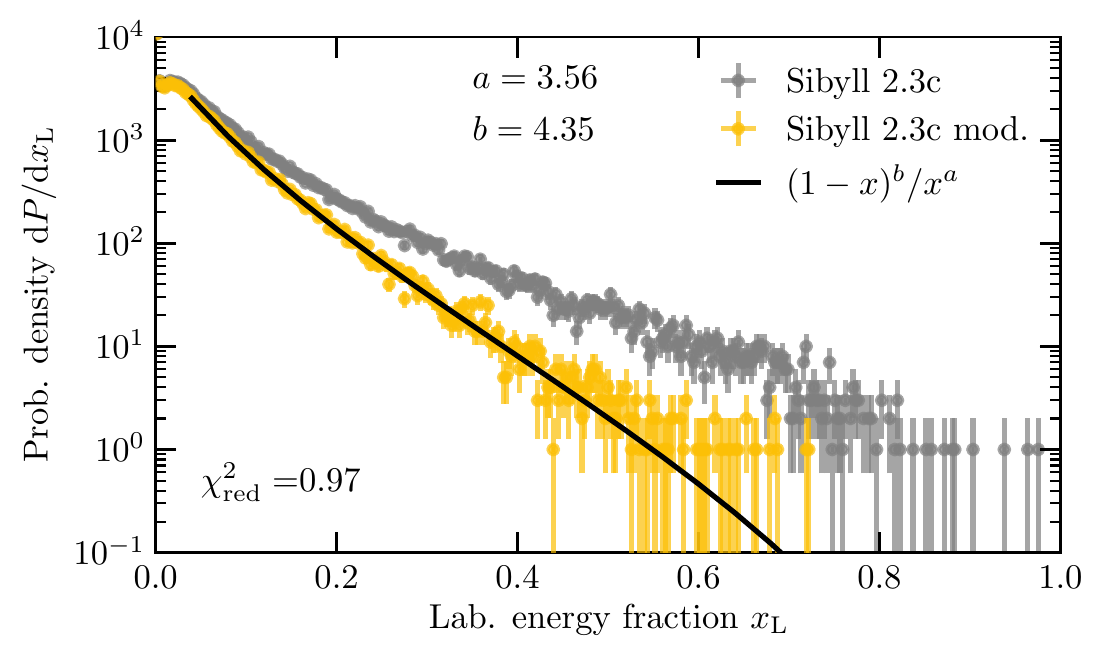}
  \vspace{0.5cm}
  \vfill
  \includegraphics[width=0.98\columnwidth,clip]{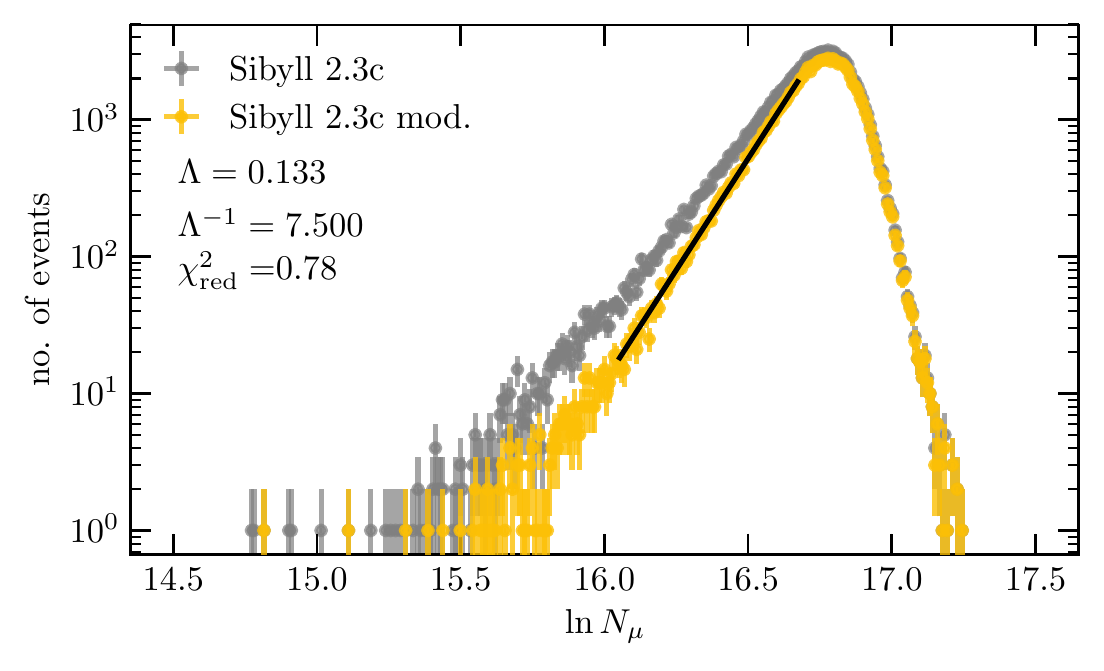}
  \caption{On the top: Inclusive production cross-section of neutral pions as a function of lab.\ energy. The nominal cross-section in \sibyll{~2.3c}~\cite{Riehn:2017mfm} is shown in gray. A modified cross-section where production of neutral pions at large $x_{\rm L}$ is suppressed is shown in yellow. On the bottom: Distribution of the number of muons at ground in case of the two production spectra shown in the figure on the top.}
  \label{fig-reshape}
\end{figure}

This connection can be verified in simulations by modifying the production cross-section and compare the resulting distribution of the number of muons. This is done in Fig.~\ref{fig-reshape}. On the top two versions of the production cross-section as a function of $x_{\rm L}~=~E_{\pi^0}/E_{\rm beam}$ are shown. In general the large-$x_{\rm L}$ behavior of the cross-section is of the form $(1-x_{\rm L})^{b}/x_{\rm L}^{a}$, where the exponent $b$ controls the suppression at large-$x_{\rm L}$ and $a$ determines the shape of the pole at zero. The two versions shown in the figure are the default cross-section in \sibyll{~2.3c}~\cite{Riehn:2017mfm} ($a=3.56$, $b=1.02$) and a modified version where the generated events have been resampled with a stronger suppression at large-$x_{\rm L}$ ($b\to 4.35$). On the bottom in Fig.~\ref{fig-reshape}, the prediction for the distribution of the number of muons at ground for the two ensembles is shown. An exponential of the form $\exp{(\ln N_{\mu}/\Lambda)}$, where $\Lambda$ is the slope parameter, is fit to the tail of the distribution. As the production cross-section of $\pi^0$ at large-$x_{\rm L}$ is reduced, the distribution of the number of muons narrows and the slope of the tail decreases ($\Lambda: 0.16 \to 0.13$).

\section{Opportunity for a measurement}
\label{sec:composition}

Having established the causal connection between the production spectrum of $\pi^0$ in the first interaction and the shape of the distribution of the number of muons of air showers, we are presented with an opportunity: if it is possible to measure the slope of the tail of the muon distribution, one effectively measures the shape of the pion production cross-section at large-$x_{\rm L}$. To test the feasibility of such a measurement we investigate the number of events that are necessary to obtain the slope of the tail of the $N_{\mu}$-distribution which is precise enough to distinguish between the predictions of the interaction models\footnote{$\delta_{\rm model}~=~ \left | \, \min(\Lambda_{\rm postLHC})-\max(\Lambda_{\rm postLHC}) \, \right | / \langle \Lambda_{\rm postLHC} \rangle ~=~ 0.2$}, $\delta_{\rm model}=0.2$.

\begin{figure}[h]
  \centering
  \includegraphics[width=\columnwidth,clip]{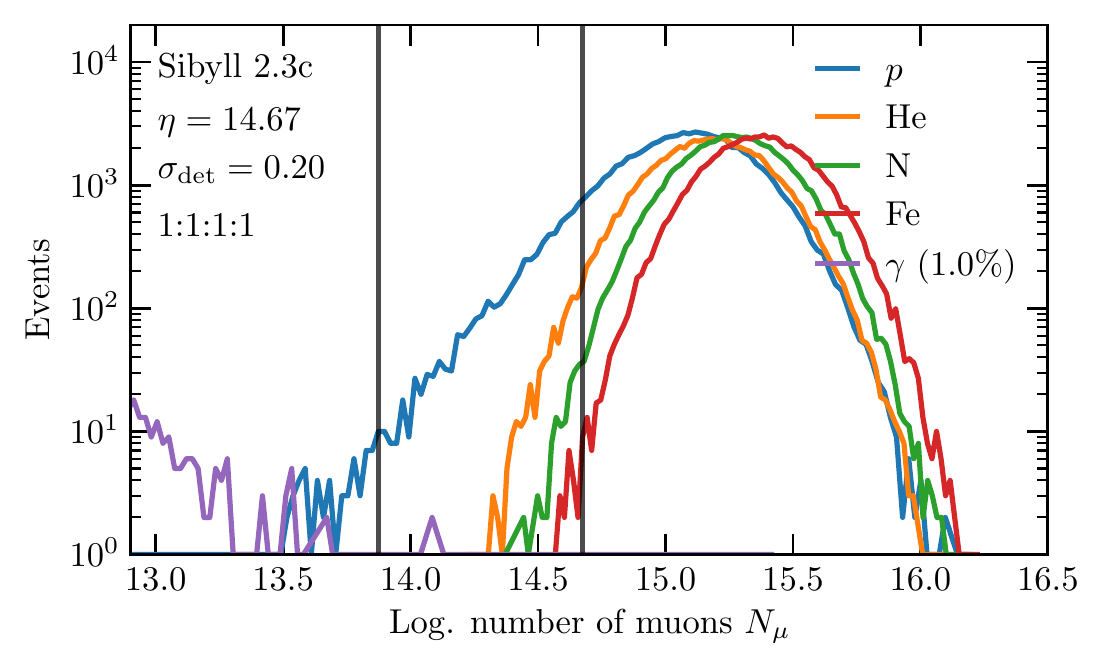}
  \caption{Distributions of $\ln N_{\mu}$ for proton, helium, nitrogen and iron induced air showers. The primary energy is $10^{19}\,$eV. Smeared with a Gaussian with relative width of 20\%.}
  \label{fig-nmu-distr-smeared}
\end{figure}

There remain three relevant factors to take into account in the measurement of $\nmu$: the resolution of the detector on the muon measurement, the resolution on the energy measurement and the mass composition of cosmic rays. The effects of finite energy resolution and finite detector resolution can be combined in a single resolution. Typical values are 15\% to 20\%~\cite{Aab:2014gua}. In order to get a clean measurement of the tail of the distribution the composition in the tail ought to be pure proton. For heavier primaries, the tail of the distribution of $\nmu$ is suppressed further and further with increasing number of nucleons the primary cosmic ray~\cite{Cazon:2018gww}. Hence the conditions for a measurement are best for proton primaries.

Much like in the case of the measurement of the interaction cross-section from the depth of the maximum of the longitudinal shower profile~\cite{Grigorov:1965aa,Auger:2012wt}, the primary elements naturally separate. This allows one to enhance the fraction of protons by selecting events with low numbers of muons, i.e.\ by going out further and further in the tail. In Fig.~\ref{fig-nmu-distr-smeared} the distributions of $\ln N_{\mu}$ of proton, helium, nitrogen and iron induced air showers are shown for the case of equal mass fractions. Resolution effects are included by adding a Gaussian smearing of 20\%. In the range $\ln N_{\mu}<14.7$ the contamination of helium in the proton sample is below 25\% and the resulting slope of the combined distribution does not differ significantly from the case of pure protons. To account for the contamination of the proton sample from the low end of muon numbers by photon initiated showers~\cite{Aab:2016agp}, the range where the slope is evaluated is limited to 0.9 units in $\ln N_{\mu}$.

\begin{figure}[h]
  \centering
  \includegraphics[width=\columnwidth,clip]{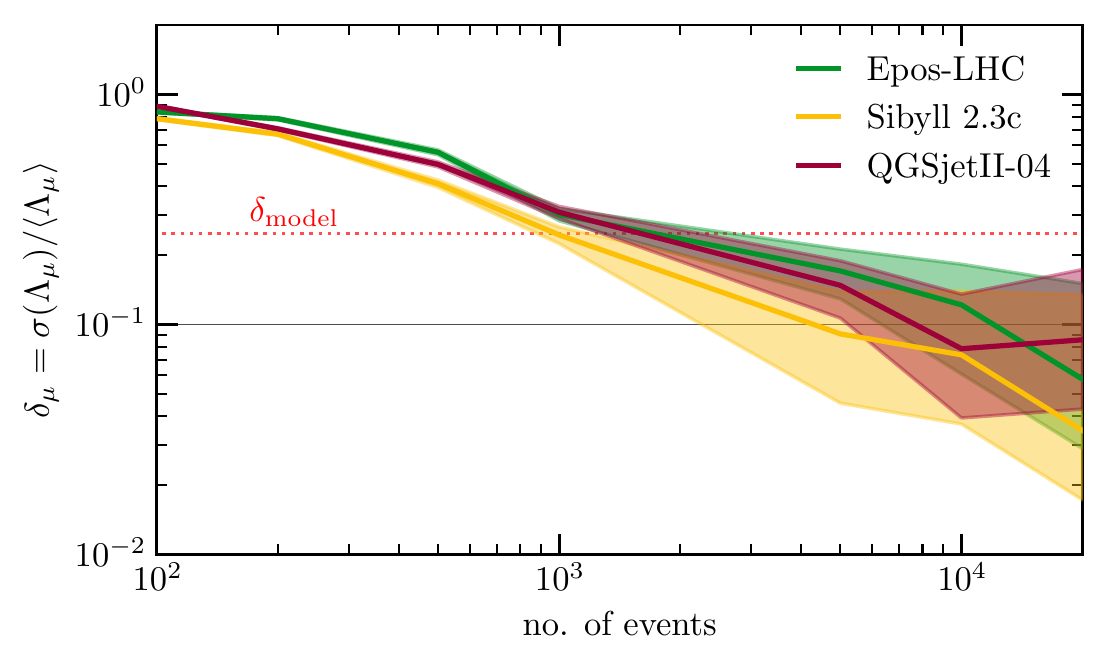}
  \caption{Expected resolution on the slope of the muon distribution as a function of the number of events. Detector resolution is 20\%. The primary composition is an equal mixture of proton, helium, nitrogen and iron. The band around the curve indicates the spread of 68\% of independent realizations of experiments with the given number of events around the median.}
  \label{fig-slope-resolution}
\end{figure}

In Fig.~\ref{fig-slope-resolution} the resolution on the slope of the muon distribution is shown as a function of the size of the statistical sample. Around 1000 events are needed to reach the desired precision. Preliminary studies show that in a case of a helium dominated mass composition ($1$:$5$:$1$:$1$), the range of low helium contamination shifts to below 14.5 and the number of events necessary to reach a precision better than $\delta_{\rm model}$ increases to $10^4$.

Given these requirements, none of the current experiments have the capabilities to perform such a measurement. However, by improving reconstruction methods~\cite{Ave:2017uiv}, extending existing detectors or deploying new experiments the measurement may become possible in the future. The upgrade of the Pierre Auger Observatory, AugerPrime~\cite{Aab:2016vlz}, for example, includes extensions at energies around $10^{17}\,$eV that may provide the required precision and event statistics~\cite{Abreu:2017vsj}. Particularly encouraging are the preliminary results on the average muon content that were obtained with the array of burried scintillators (AMIGA)~\cite{AMIGA} with around a year of data~\cite{UHECRAMIGA}.

Through the combined measurement of the surface and radio signal of inclined air showers in the energy range from $10^{16.5}\,$eV to $10^{18}\,$eV, GRANDProto300, a pathfinder experiment for the proposed ``Giant Radio Array for Neutrino Detection'' (GRAND) experiment~\cite{Alvarez-Muniz:2018bhp}, may provide another good opportunity. The high duty cycle of surface detector stations and radio antennas provides the statistical power, while the focus on inclined showers (a pre-requisite for radio detection with sparse antenna arrays) insures a clean muon signal in the surface detectors. Although the feasibility of the measurement of the radio emission from extensive air showers on such a scale has not been demonstrated before, energy resolutions of 15\% to 20\% have been achieved~\cite{Alvarez-Muniz:2018bhp,Aab:2016eeq}.

\section{Impact} 
The benefits of having a detailed measurement of the shape of the distribution of the number of muons in proton induced extensive air showers are twofold. First and most immediate, it would constrain hadronic interaction models and help finding a solution to the muon problem. As the width of the muon distribution is based on the energy flow in the first interaction, the measurement is sensitive to exotic phenomena that would affect this flow. Scenarios invoking the violation of Lorentz invariance~\cite{Coleman:1998ti}, for example, predict that beyond some energy threshold the decay of particles, including neutral pions, is suppressed. This then changes the shape of the distribution of hadronic energy, the tail and width of the muon distribution changes.

Second, the measurement of the pion production cross-section at large-$x_{\rm L}$ in extensive air showers nicely complements the measurements by LHCf at the LHC~\cite{Adriani:2015iwv}. In terms of fundamental physics, the production cross-sections at large-$x_{\rm L}$ are sensitive to the momentum distribution of valence quarks in the projectile and the very small-$x$ behavior of the gluon distributions in the target. The predictions on the nature of particle production under such extreme conditions, range from the formation of a ``color glass condensate''~\cite{McLerran:1993ni}, to complete absorption of valence quarks and disappearance of leading particles~\cite{Drescher:2005ak}.

\section{Conclusion}

In this study we have shown that, the forward production cross-section of neutral pions can be accessed experimentally through the measurement of the tail of the shower-to-shower distribution of the number of muons in proton induced extensive air showers.

%
\bibliography{local}

\end{document}